\documentclass[prb,twocolumn]{revtex4}
\usepackage{graphicx}
\usepackage{amsmath}
\usepackage{amsbsy}
\usepackage{bm}

\newcommand{\notop}{{{}}}

\begin{document}

\title{Non-collinear magnetoconductance of a quantum dot}
\author{Jonas N. Pedersen$^{1,2}$, Jesper Q. Thomassen$^1$, and Karsten
Flensberg$^1$} \affiliation{Nano-Science Center, Niels Bohr
Institute, Universitetsparken 5, 2100
Copenhagen, Denmark.\\
$^2$Department of Physics, University of Lund, Box 118, 22100 Lund, Sweden.}
\date{\today}

\begin{abstract}
We study theoretically the linear conductance of a quantum dot
connected to ferromagnetic leads. The dot level is split due to a
non-collinear magnetic field or intrinsic magnetization. The
system is studied in the non-interacting approximation, where an
exact solution is given, and, furthermore, with Coulomb
correlations in the weak tunneling limit. For the non-interacting
case, we find an anti-resonance for a particular direction of the
applied field, non-collinear to the parallel magnetization
directions of the leads. The anti-resonance is destroyed by the
correlations, giving rise to an interaction induced enhancement of
the conductance. The angular dependence of the conductance is thus
distinctly different for the interacting and non-interacting cases
when the magnetizations of the leads are parallel. However, for
anti-parallel lead magnetizations the interactions do not alter
the angle dependence significantly.
\end{abstract}

\pacs{72.25.-b,85.75.-d,73.23.Hk,73.63.-b}
\maketitle

\section{Introduction}

Spin transport through hybrid structures of both semiconductor and
metal systems is now a well-established field of
research\cite{zutic04}. In recent years, also spin transport
through quantum dots formed by nanotubes, molecular systems or
nanoparticles has started to attract attention both
experimentally\cite{tsuk99,orga01,hoff03,zhao,kim02,pett04,nygard}
and
theoretically.\cite{rudz01,konig03,mart03a,mart03b,braun04,braig04,rudz04,frans05}
One of the motivations for this has been the possibility of using
quantum dots with a single spin as qubits, but naturally also a
number of fundamental questions arise on how the interplay of spin
coherence and interactions influences the transport.

When the spin coherence is maintained during the electron passage
through the mesoscopic structure, the transport needs to be
described in a coherent fashion at least for spin degree of
freedom. Even for weak tunneling one therefore has to modify the
usual rate equation approach if the magnetization directions are
non-collinear.\cite{konig03,braun04,braig04} For stronger
tunneling coupling also many of the non-equilibrium Green's
function methods fail to describe the case when both interactions
and spin coherence are important, such as for example the Hubbard
I approximations used in
Refs.~\onlinecite{rudz04,frans05}.\cite{novotny}

For the strong coupling case, a particular interesting situation
occurs when the mesoscopic system has a magnetization which is
non-collinear with respect to that of the ferromagnetic leads,
which is what is studied in this paper. We study the conductance
of a single-level quantum dot connected to ferromagnetic leads,
and, in addition, the spin states of the quantum dot being split
due to, e.g., an applied magnetic field. Experimentally, one could
realize this geometry if the magnetic leads are fabricated as
magnetic thin films, where the magnetic field is strongly pinned
in the plane parallel to the film, so that a small magnetic field
component can be applied to the quantum dot without changing the
magnetization direction of the leads. The geometry of the device
is illustrated in Fig.~1.
\begin{figure}[tbp]
\includegraphics[width=.3\textwidth]{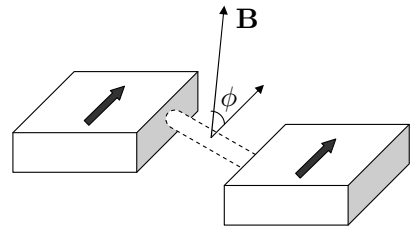}
\caption{Schematic drawing of the quantum dot system, connected to two
ferromagnetic leads. In addition, there is an applied magnetic which spin
polarizes the quantum dot in a direction non-collinear to the magnetization
of the leads. In this paper, we focus on how the conductance depends on the
angle $\protect\phi$, and the interplay between interactions and spin
interference.}
\label{fig:model}
\end{figure}
We describe the system using a modified Anderson Hamiltonian where
the leads are assumed to be polarized and the magnetic field gives
rise to an energy splitting of the dot level. This gives an
effective double-slit geometry with two different paths through
the central region. In the case of parallel magnetizations of the
leads and no Coulomb repulsion on the dot, we find sharp
anti-resonances in the conductance for certain angles $\phi$,
which is explained as destructive interference. If the Coulomb
repulsion on the dot is increased, a cross-over to a simple
spin-valve behavior is seen. Changing the angle $\phi$ in an
experimental setup can therefore indicate whether interactions on
the dot are important or if a single-electron picture is
sufficient. On the other hand, in the case of anti-parallel
magnetizations of the leads no change in the angular dependence is
observed when the Coulomb repulsion on the dot is increased.
Finally, we limit ourselves in this paper to the study of linear
response, so non-equilibrium effects associated with spin
accumulation are not included.

The paper is organized as follows. In Section~\ref{sec:model}, we
set up the model Hamiltonian. We start by calculating the
conductance for the non-interacting case in
Section~\ref{sec:nonint}, and in Section~\ref{sec:cotun} we
consider the cotunneling limit. Finally, in Section~\ref{sec:disc}
we discuss the result and its limitations.

\section{Model Hamiltonian}

\label{sec:model}

The model Hamiltonian of the quantum dot coupled to magnetic leads is
\begin{equation}  \label{Hmodel}
H=H_{LR}^{{}}+ H_T + H_D,
\end{equation}
where
\begin{equation}  \label{HLR}
H_{LR}^{{}}=\sum_{\alpha=L,R,k\sigma} \xi_{\alpha,k\sigma}^{{}}
c^\dagger_{\alpha,k\sigma} c^{{}}_{\alpha,k\sigma},
\end{equation}
where $\alpha$ denotes the left or right electrode and where we
have assumed that the polarizations of the electrodes are parallel
or anti-parallel. The magnetization of the quantum dot is not
necessarily parallel to those of the leads and in the spin basis
of the leads, the dot Hamiltonian is
\begin{equation}  \label{HD}
H_{D}^{{}}=\sum_{\sigma} \xi_{0}^{{}}d^\dagger_\sigma
d^{{}}_\sigma
+Un_\uparrow^{{}}n_\downarrow^{{}}+\sum_{\sigma\sigma^{\prime}}\,\mathbf{B}
\cdot\mathbf{\sigma}_{\sigma\sigma^{\prime}}^{{}}
d^\dagger_\sigma d^{{}}_{\sigma^{\prime}},
\end{equation}
where $\xi_0$ is the orbital quantum dot energy, $B$ represents the magnetic
field splitting, and $U$ is the Coulomb energy for double occupancy. In a
diagonal basis, the dot Hamiltonian is
\begin{equation}  \label{HDd}
H_{D}^{{}}=\sum_{\sigma} (\xi_{0}^{{}}+\sigma B)\tilde{d}^\dagger_\sigma
\tilde{d}^{{}}_\sigma +U\tilde{n}_\uparrow^{{}}\tilde{n}_\downarrow^{{}},
\end{equation}
where the $d$ and $\tilde{d}$ operators are related by the unitary rotation
\begin{equation}  \label{dU}
d_\sigma=\sum_{\sigma^{\prime}} R_{\sigma\sigma^{\prime}}^{{}}
\tilde{d}_{\sigma^{\prime}},\quad \mathbf{R}=\left(
\begin{array}{cc}
\cos(\phi/2) & \sin(\phi/2) \\
-\sin(\phi/2) & \cos(\phi/2)
\end{array}
\right).
\end{equation}
Finally, the tunneling Hamiltonian is
\begin{equation}  \label{HT}
H_T^{{}}=\sum_{\alpha=L,R}\sum_{k\sigma\sigma^{\prime}} \left(
t_{\alpha,k\sigma}^{{}}R^{{}}_{\sigma\sigma^{\prime}}
c^\dagger_{\alpha,k\sigma}\tilde{d}^{{}}_{\sigma^{\prime}}+\mathrm{h.c.}
\right).
\end{equation}
Here we allow for the tunneling matrix element to be spin
dependent, because the states in the leads depend on spin
direction. The possibility of spin flip processes during the
tunneling process is not included explicitly here, however, it
could easily be included by replacing the diagonal matrix
$t_\sigma$ by a non-diagonal tunneling matrix. Depending on
parameters this would correspond to having an angle between the
lead magnetizations, which would modify the details but not the
general behaviors that we discuss.

\section{Resonant tunneling without correlations}

\label{sec:nonint}

We start by discussing the current through the system in absence
of correlations, i.e. $U=0$. This treatment is relevant where
mean-field approaches are valid (of course $\xi_0^{{}}$ should be
determined self-consistently), e.g. for a resonant tunneling
junction system. The conductance follows from the Landauer formula
as
\begin{equation}  \label{I0}
G=\frac{e^2}{h} \int d\epsilon\, T(\epsilon) \left(-\frac{\partial
n_\mathrm{F}(\epsilon)}{\partial \epsilon}\right),
\end{equation}
where $T(\epsilon)$ is the total transmission coefficient for both
spin directions, and $n_\mathrm{F}$ is the Fermi-Dirac
distribution function. In terms of the retarded and advanced
Green's function the transmission coefficient is\cite{meir92}
\begin{equation}  \label{T0}
T(\epsilon)=\mathrm{Tr}\left[\mathbf{G}^a(\epsilon)
\mathbf{\Gamma}^R(\epsilon)\mathbf{G}^r(\epsilon)
\mathbf{\Gamma}^L(\epsilon)\right],
\end{equation}
with (using the spin basis where $H_D$ is diagonal)
\begin{subequations}
\begin{align}  \label{Gr0}
\mathbf{G}^{r,a}(\epsilon)&=\left(\mathbf{G}_0^{-1}-\mathbf{\Sigma}^{r,a}_L
(\epsilon)-\mathbf{\Sigma}^{r,a}_R(\epsilon)\right)^{-1}, \\[3mm]
[\mathbf{G}_0(\epsilon)]_{\sigma\sigma^{\prime}}&
=\delta_{\sigma,\sigma^{\prime}}(\epsilon-\xi_0-\sigma B)^{-1}, \\[3mm]
[\mathbf{\Sigma}^{r,a}_\alpha
(\epsilon)]_{\sigma\sigma^{\prime}}&=\sum_{k\sigma^{\prime\prime}}
(R^\dagger)_{\sigma\sigma^{\prime\prime}}^{{}}
|t^{{}}_{\alpha,k\sigma^{\prime\prime}}|^2 g^{r,a}_{\alpha,
k\sigma^{\prime\prime}}(\epsilon)R_{\sigma^{\prime\prime}
\sigma^{\prime}}^{{}}\,,
\end{align}
where $g^{r,a}_{\alpha,
k\sigma}(\epsilon)=(\epsilon-\xi_{\alpha,k\sigma}^{{}}\pm
i0^+)^{-1}$ and
$\mathbf{\Gamma}_{\alpha}=-i\left[\mathbf{\Sigma}^{r}_\alpha-\mathbf{\Sigma}
^{a}_\alpha\right]$. In order to simplify the analysis, we assume
that parts corresponding to the real parts of $g^{r,a}$ give
constant shift of $\xi_0$ and hence can be incorporated in
$\mathbf{G}_0$, and, furthermore, that the tunnel width functions
$\mathbf{\Gamma}_{L,R}$ are constant in energy. With these
assumptions
\end{subequations}
\begin{align}  \label{GammaLRphi}
\mathbf{\Gamma}_{L,R}(\epsilon)=\frac{\Gamma_{L,R}^0}{2}
\left(\begin{array}{cc}
1+P_{L,R}^{{}} \cos\phi & P_{L,R}^{{}} \sin\phi \\
P_{L,R}^{{}} \sin\phi & 1-P_{L,R}^{{}} \cos\phi
\end{array}
\right),
\end{align}
where
\begin{equation}  \label{Gammadef}
\Gamma_{\alpha}^0=2\pi\sum_{k\sigma}|t^{{}}_{\alpha,k\sigma}|^2\delta(
\varepsilon-
\xi_{\alpha,k\sigma}^{{}})=\sum_\sigma\Gamma_{\alpha,\sigma}^0,
\end{equation}
and $P_\alpha$ denotes the polarization of the tunneling from lead $\alpha$
defined through $\Gamma^0_{\alpha,\sigma}=\frac{1}{2}\left(1+\sigma
P_\alpha^{{}}\right)\Gamma^0_\alpha$. Finally, we have
\begin{align}  \label{Gr0model}
\mathbf{G}^{r,a}(\epsilon)=\left([\mathbf{G}^{r,a}_0(\epsilon)]^{-1}\pm
i( \mathbf{\Gamma}_L+ \mathbf{\Gamma}_R)/2\right)^{-1}.
\end{align}
With these formulae it is straightforward to find the transmission
coefficient.

In the following, we study the current using the simplifying case of fully
polarized leads, i.e. half-metallic contacts. For the parallel
configuration, $P_L=P_R=1$, we obtain for the transmission function
\begin{align}  \label{TP}
T_P^{{}}(\epsilon)&=\frac{\Gamma_L^0\Gamma_R^0}{B^2}\frac{[x-\cos\phi]^2}
{(1-x^2)^2+y^2(x-\cos\phi)^2},
\end{align}
where
\begin{equation}  \label{xy}
x=\frac{\epsilon-\xi_0}{B},\quad
y=\frac{\Gamma_L^0+\Gamma_R^0}{2B}=\frac{\Gamma^0}{2B}.
\end{equation}
\begin{figure}[tbp]
\includegraphics[width=.475\textwidth]{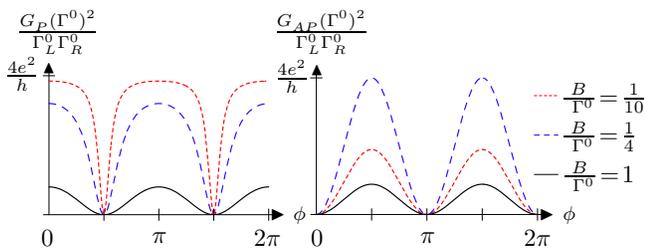}
\caption{(Colors online.) The conductance for the non-interacting
case in both the parallel (left panel) and the anti-parallel
(right panel) cases for different values of the magnetic strength
and at the symmetric point $\protect\xi_0=0$. For the parallel
case, destructive interference leads to an anti-resonance for
$\protect\phi=\protect\pi/2$. The anti-resonance sharpens as
$B/\Gamma^0$ decreases. The conductance is maximal when the
applied field is parallel to magnetizations in the leads
($\protect\phi=0$ or $\protect\pi$). In contrast, for the
anti-parallel case the current is maximal for $\protect\phi
=\protect\pi/2$ and vanishes at $\protect\phi=0$ and $\protect\pi$
because of the spin-valve effect. } \label{fig:Gexact}
\end{figure}

Let us look at the symmetric condition $x=0$, where the incoming
energy lies symmetrically between the spin split levels:
\begin{align}  \label{TP2}
T_P(\xi_0)=\frac{\Gamma_L^0\Gamma_R^0}{B^2}\frac{\cos^2\phi}
{1+y^2\cos^2\phi}.
\end{align}
This yields a sharp anti-resonance when the dot magnetization is
perpendicular to the magnetization direction of the leads. The
anti-resonance occurs due to a destructive interference between
the transmissions through the two spin split levels, as explained
in the following section. It is interesting to note that the zero
transmission at $\phi=\pi/2$ is not broadened by higher order
tunneling events, as one might have expected. Moreover, for the
non-symmetric situation $x\neq 0$, the anti-resonance remains, but
is shifted away from $\phi=\pi/2$ to the point where $x=\cos\phi$,
which follows from Eq.~\eqref{TP}. The anti-resonance found here
is in fact similar to an effect exploited in single atom
transmission through optical lattices.\cite{mich04}

Next, we look at the antiparallel situation. Therefore we set
$P_L=-P_R=1$ and find
\begin{align} \label{TAP}
T_{AP}(\epsilon)=\frac{4\Gamma_L^0\Gamma_R^0}{B^2}\frac{\sin^2\phi}{D},
\end{align}
where
\begin{align}
D &= 4(1-x^2)^2+2y_L^{{}} y_R^\notop+y_L^2y_R^2/4+(y_R^2+y_L^2)x^2
\notag
\\
&\quad+2(y_L^2-y_R^2)x\cos\phi +(y_L^\notop-y_R^\notop)^2\cos^2\phi,
\label{D}
\end{align}
with $y_\alpha=\Gamma_\alpha^0/B$. Note that for the symmetric
coupling $\Gamma_L^0=\Gamma_R^0$, the only angle dependence is the
$\sin^2\phi$ factor in the numerator. In Fig.~\ref{fig:Gexact} we
show examples of the angle dependence of the conductance for both
the parallel and antiparallel cases.

A difference between the parallel and antiparallel geometry is the
existence of an optimal $B$-value in the antiparallel case.
Considering the symmetric condition $x=0$, we find for the
parallel case ($\phi=0,\pi$)
\begin{align}
T_P^{\mathrm{
max}}(\xi_0)=\frac{\Gamma_L^0\Gamma_R^0}{B^2+(\Gamma^0/2)^2},
\end{align}
which decreases when increasing $B$, because the levels are moved
away from the chemical potentials of the leads. For the
antiparallel geometry, the maximal transmission at the symmetric
condition is ($\phi=\pi/2,~3\pi/2$)
\begin{align}
T_{AP}^{\mathrm{max}}
(\xi_0)=\frac{\Gamma_L^0\Gamma_R^0}{\left[B+\Gamma^0_L\Gamma_R^0/(4B)\right]^2},
\end{align}
and the optimal value $B_{\mathrm{
opt}}=\sqrt{\Gamma^0_L\Gamma_R^0/4}$ is found. This value is a
comprise between the spin blockade for $B=0$ and the levels being
far from the chemical potentials of the leads for $B/\Gamma^0\gg
1$.

\begin{figure}[tbp]
\includegraphics[width=.35\textwidth]{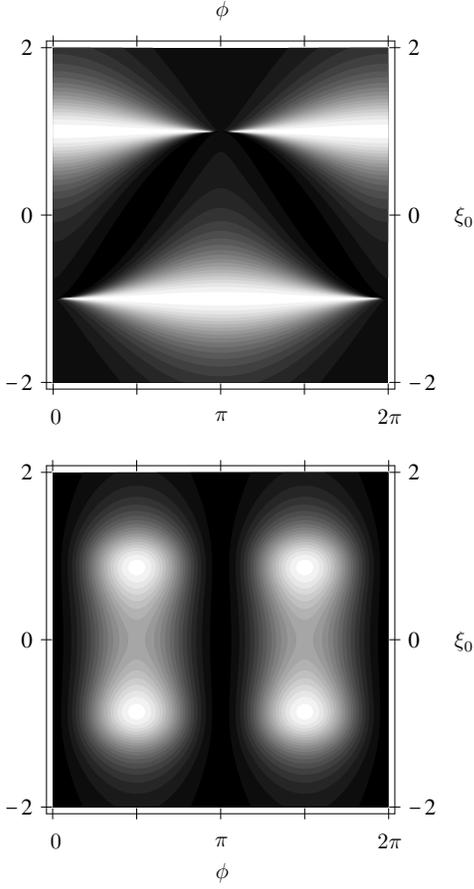}
\caption{Contour plots of the conductance in the non-interacting
case for the parallel (top panel) and antiparallel (lower panel)
geometry for different values of the angle $\phi$ and the orbital
dot quantum dot energy $\xi_0^{{}}/\Gamma_0^{{}}$
($B/\Gamma_0^{{}}=1$ in both figures). For the antiparallel
geometry the same angular dependence is observed for all
$\xi_0^{{}}$, whereas a change is seen for the parallel geometry
as explained in the text. } \label{fig:contour}
\end{figure}

Finally, Fig.~\ref{fig:contour} shows how the conductance changes
when $\xi_0^\notop$ is varied. In the parallel case, the
destructive interference only occurs close to the symmetric
condition, $x\approx 0$, whereas when both levels are above or
below the resonance a simple spin valve effect is observed. If one
of the levels is exactly at the chemical potentials of the leads
the conductance is constant ($\xi_0^{{}}=\pm B$). For the
antiparallel case no qualitative change in the angular dependence
is seen.

\section{Cotunneling current for the interacting case}

\label{sec:cotun}

In the presence of interactions the interesting interference
effect discussed above can be significantly changed. Naturally,
this is in general a difficult problem to tackle. Here we look at
the case where the position of the levels is far from resonance as
defined below, and a second order perturbation calculation
suffices. This is the ``cotunneling" regime, where an electron is
transferred through the system in a two-electron process. The
cotunneling appproach is valid far from resonance, i.e. when
$|\varepsilon_\uparrow^{{}}-\mu|$ and
$|\varepsilon_\downarrow^{{}}+U|$ are larger than the bias,
temperature and tunneling broadening.

For the resonant case the cotunneling result diverges and there is
a cross-over to the sequential tunneling regime, see
App.~\ref{sec:seq}. In these points interactions are not
important, because transport is determined by only one level.

We calculate the current using a scattering formalism and the
transition rate per unit time between an initial state $i$ (with
energy $E_i$) and a final state $f$ (with energy $E_f$) is
calculated as
\begin{align}  \label{Gamma}
\Gamma_{fi}=\frac{2\pi}{\hbar} |\langle
f|T|i\rangle|^2\delta(E_f^\notop-E_i^\notop),
\end{align}
where the transition operator $\hat{T}$ is
\begin{align}  \label{Toperator}
\hat{T}=H_T+H_T\frac{1}{E_i^\notop-H_0^\notop+i\eta}\hat{T}
\end{align}
and $H_0=H_{LR}^\notop+H_D^\notop$. The current is now given by
processes which bring an electron from the left to the right lead,
$\Gamma_{RL}^\notop$, minus the opposite processes:
\begin{align}
J=e(\Gamma_{RL}^\notop-\Gamma_{LR}^\notop).
\end{align}
The transition rates will only be calculated to lowest
non-vanishing order in the coupling. Thus, in the initial state
the leads can be considered as two non-interacting electron gasses
(here assumed fully spin-polarized) with different chemical
potentials. The probability to find the lead $\alpha$ in a state
$\nu_\alpha^\notop$ is denoted $W_{\nu_{\alpha}}$. Because the
leads are non-interacting, $W_{\nu_{\alpha}}$ satisfies
\begin{align}
&\sum_{\nu_{\alpha}}W_{\nu_{\alpha}^\notop}\langle\nu_{\alpha}^\notop|
c^\dag_{\alpha,k\sigma}
c_{\alpha,k\sigma}^\notop|\nu_\alpha^\notop\rangle
=n_\mathrm{F}^\notop(\varepsilon_{\alpha,k\sigma}^\notop-\mu_\alpha^{{}}).
\end{align}
\begin{figure}
\includegraphics[width=.35\textwidth]{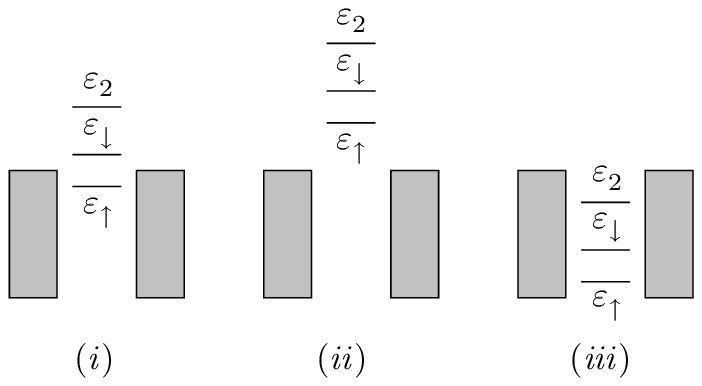}
\caption{The three different cotunneling regimes considered
($\varepsilon_2^{{}}
=\varepsilon_\uparrow^{{}}+\varepsilon_\downarrow^{{}}+U$).}
\label{fig:CoRegimes}
\end{figure}

Three different regimes are considered (see Fig.~\ref{fig:CoRegimes}):\\
\textit{i}) the level with the lowest energy is always occupied
and the other energy level always empty, \textit{i.e.}
$\varepsilon_\uparrow^{{}}<\mu$ and $\varepsilon_\downarrow^{{}}
>\mu$,\\
\textit{ii}) both levels are empty, i.e.
$\varepsilon_\uparrow^{{}}>\mu$ and $\varepsilon_\downarrow^{{}}
>\mu$,\\
\textit{iii}) both levels occupied, i.e.
$\varepsilon_\uparrow^{{}}<\mu$ and $\varepsilon_\downarrow^{{}}+U
<\mu$.\\
We show how the result is obtained for the first regime with
parallel geometry and only state the results for the other
regimes.

\begin{figure}[tbp]
\includegraphics[width=.3\textwidth]{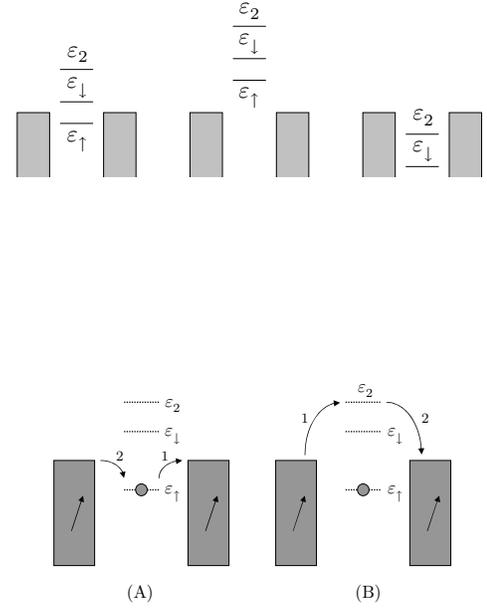}
\caption{The two possible second order processes contributing to
the tunneling amplitude, here shown for the parallel configuration
in the first regime. Because of interactions, the path to the
right is suppressed when the lower energy spin-state is occupied
by one electron. Hence the interactions tend to destroy the
interference effects seen in Fig.~\protect\ref{fig:Gexact}, valid
for $U=0$, and instead one gets the angular dependence shown in
Fig.~\protect\ref{fig:Gcotun}.} \label{fig:cotun}
\end{figure}

In case \textit{i}), the initial states can be written as\\
$|i\rangle=|\nu_L^\notop,\nu_R^\notop,\uparrow\rangle$, with
energy $E_i^{{}}=E^\notop_{\nu_L}
+E^\notop_{\nu_R^\notop}+\varepsilon_\uparrow^\notop$ and
probability $W_{\nu_L^\notop} W_{\nu_R^\notop}$. First, we outline
the derivation for $\Gamma_{RL}^\notop$ in the parallel
configuration. In the fully polarized case, the leads contain only
electrons with one kind of spin, say spin-up, and the final states
are $|f_{kk^{\prime}}\rangle~=~c_{R,k\uparrow}^\dag
c_{L,k^{\prime}\uparrow}^\notop|i\rangle$, meaning that an
electron is moved from the left to the right lead in the process.
Inserting the expressions for the final states and the tunneling
Hamiltonian in Eqs.~(\ref{Gamma}) and (\ref{Toperator}), we see
that the first non-vanishing term is second order in the coupling
$H_T^\notop$. Moreover, two processes give a contribution: A)
first the electron on the dot jumps to the right lead and later an
electron from the left lead refills the dot, or B) a left lead
electron enters the dot, creating a double occupied state, and
later leaves the dot to the right. The two processes are sketched
in Fig.~\ref{fig:cotun}.
The scattering rate $\Gamma_{RL}^{(i)}$ can now be written as
\begin{align}
\Gamma_{RL}^{(i)}=\frac{2\pi}{\hbar}\!\!\sum_{kk^{\prime}\nu_L^\notop\nu_R^
\notop}W_{\nu_L^\notop}W_{\nu_R^\notop}
\left|M_A^\notop+M_B^\notop\right|^2\delta(\varepsilon_{R,k\uparrow}^\notop
-\varepsilon_{L,k^{\prime}\uparrow}^\notop),
\end{align}
where
\begin{align}
M_A^\notop&=\sum_{k_1^\notop k_2^\notop}t^*_{L,k_2^\notop\uparrow}
t^\notop_{R,k_1^\notop\uparrow}|R^\notop_{\uparrow,\uparrow}|^2 \\
&\quad\times\langle f_{kk^{\prime}}^\notop|\tilde{d}_\uparrow^\dag
c_{L,k_2\uparrow}^\notop\frac{1}{E^{{}}_i-H_0^\notop}c^\dag_{R,k_1^\notop
\uparrow}\tilde{d}_\uparrow^\notop |i\rangle,  \notag \\
M_B^\notop&=\sum_{k_1^\notop k_2^\notop}t^{*}_{L,k_2^\notop\uparrow}
t^{{}}_{R,k_1^\notop\uparrow} |R^\notop_{\downarrow,\uparrow}|^2 \\
&\quad\times\langle
f_{kk^{\prime}}^\notop|c^\dag_{R,k_1^\notop\uparrow}
\tilde{d}^\notop_\downarrow
\frac{1}{E^{{}}_i-H_0^\notop}\tilde{d}_\downarrow^\dag
c^\notop_{L,k_2^\notop\uparrow} |i\rangle.  \notag
\end{align}
The matrix elements $M_A^\notop$ and $M_B^\notop$ are easily calculated, and
after inserting the coupling constants in Eq.~\eqref{Gammadef} and assuming
them to be independent of energy, we arrive at
\begin{align}
\Gamma^{(i)}_{RL}&=\frac{\Gamma^0_L\Gamma^0_R}{2\pi}\int d\epsilon
\left[\frac{|R_{\uparrow,\uparrow}^\notop|^2}{\varepsilon^\notop_\uparrow-\epsilon}
-\frac{|R_{\downarrow,\uparrow}^\notop|^2}
{\epsilon-\varepsilon_\downarrow^\notop-U}\right]^2  \notag \\
&\qquad\times
n_\mathrm{F}^\notop(\epsilon\!-\!\mu_L^{{}})[1-n_\mathrm{F}^\notop(\epsilon\!-\!\mu_R^{{}})].
\end{align}
For $\Gamma^{(i)}_{LR}$ exactly the same calculations are carried
out and we finally obtain for the current
\begin{align}
J_P^{(i)}&=\frac{e}{\hbar}\frac{\Gamma^0_L\Gamma^0_R}{2\pi}\int
d\epsilon
\left[\frac{|R_{\uparrow,\uparrow}^\notop|^2}{\varepsilon^\notop_\uparrow-\epsilon}
-\frac{|R_{\downarrow,\uparrow}^\notop|^2}{\epsilon-\varepsilon_\downarrow^\notop-U}\right]^2  \notag \\
&\qquad\times
[n_\mathrm{F}^\notop(\epsilon\!-\!\mu_L^{{}})-n_\mathrm{F}^\notop(\epsilon\!-\!\mu_R^{{}})].
\end{align}
At small bias voltage, we use that $|\varepsilon|
\ll|\varepsilon^\notop_\uparrow|,|\varepsilon^\notop_\downarrow+U|$ and when
inserting the form of $\mathbf{R}$ from Eq.~\eqref{dU}, we obtain
\begin{align}  \label{Jparr}
J_P^{(i)}=e^2V\frac{\Gamma^0_L\Gamma^0_R}{h}\left[\frac{\cos^2(\phi/2)}
{\xi_0^\notop-B}
+\frac{\sin^2(\phi/2)}{\xi_0^\notop+B+U}\right]^2,
\end{align}
where $\mu_L^{{}}-\mu_R^{{}}=eV$. Note that the two terms in the
square brackets are the amplitudes for the two tunneling processes
shown in Fig.~\ref{fig:cotun}. In Eq.~\eqref{Jparr} it is evident
that a destructive interference between the two paths can occur at
certain angles.

Of course, a similar calculations can be performed for the
antiparallel configuration, and with the same constraints the
result is
\begin{align}
J_{AP}^{(i)}&=e^2V\frac{\Gamma^0_L\Gamma^0_R}{4h} \left[\frac{1}
{\xi_0^\notop-B} -\frac{1}{\xi_0^\notop+B+U}\right]^2\sin^2\phi .
\label{JAparr}
\end{align}
Looking at this expression, it is clear that the constructive
interference for the anti-parallel case is not sensitive to the
angle, because both paths have the angular dependence.

\begin{figure}[tbp]
\includegraphics[width=.485\textwidth]{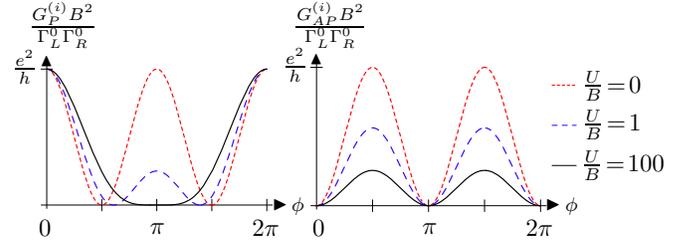}
\caption{(Colors online) The conductance for the
\textit{interacting} case in both the parallel (left panel) and
the anti-parallel (right panel) cases for different values of the
Coulomb interaction and for the symmetric point $\protect\xi_0=0$.
For the parallel case, we see how the anti-resonance at
$\protect\phi=\protect\pi/2$ is destroyed by interactions, due to
the blocking of one of the paths in Fig.~\protect\ref{fig:cotun}.
For large $U$ one instead has a cross-over to a spin-valve effect
at $\protect\phi=\protect\pi$, where the dot is occupied by an
electron with spin opposite to the lead electrons. The angle
dependence thus dramatically depends on the interaction. This is
not so for the anti-parallel case shown in the right panel, where
merely an overall suppressing is seen. } \label{fig:Gcotun}
\end{figure}

The cotunneling results for regime \textit{i}) are plotted in
Fig.~\ref{fig:Gcotun} for different values of the Coulomb
repulsion $U$. It is easily verified that to second order in
$\Gamma$ and $U=0$ the cotunneling results \eqref{Jparr} and
\eqref{JAparr} and the exact results for $U=0$ derived above
coincide.

In the other two regimes similar calculations give
\begin{align}
J_{P}^{(ii)}&=e^2V\frac{\Gamma^0_L\Gamma^0_R}{h}
\left[\frac{\cos^2(\phi/2)}{
\xi_0^\notop-B}+\frac{\sin^2(\phi/2)}{\xi_0^\notop+B}\right]^2,  \\
J_{AP}^{(ii)}&=e^2V\frac{\Gamma^0_L\Gamma^0_R}{4h}
\left[\frac{1}{\xi_0^\notop+B}-\frac{1}{\xi_0^\notop-B}\right]^2\sin^2\phi
\end{align}
for both levels being empty, and
\begin{align}
J_{P}^{(iii)}&=e^2V\frac{\Gamma^0_L\Gamma^0_R}{h}
\left[\frac{\cos^2(\phi/2)}
{\xi_0^\notop-B+U}+\frac{\sin^2(\phi/2)}{\xi_0^\notop+B+U}\right]^2,  \\
J_{AP}^{(iii)}&=e^2V\frac{\Gamma^0_L\Gamma^0_R}{4h}
\left[\frac{\sin\phi}{\xi_0^\notop+B+U}-\frac{\sin\phi}{\xi_0^\notop-B+U}\right]^2,
\end{align}
for both levels being occupied. Note that for $U=0$ holds
$J^{(i)}=J^{(ii)}=J^{(iii)}$, and it agrees with the
non-interacting results in Sec.~\ref{sec:nonint} to lowest order
in $\Gamma_L^0\Gamma_R^0$.

\begin{figure}[tbp]
\includegraphics[width=.4\textwidth]{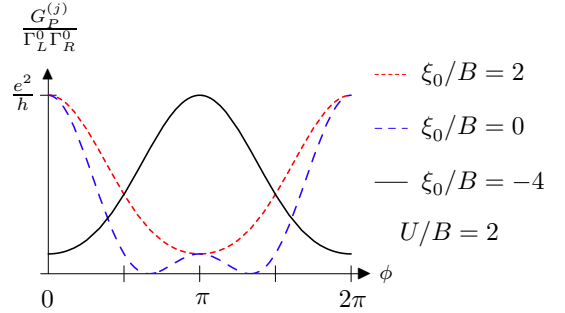}
\caption{(Colors online.) The cotunneling result for the parallel
geometry in the three different regimes [$j=i,~ii,~iii$]  with
\textit{i}) $\xi_0^{{}}=0$, \textit{ii}) $\xi_0^{{}}/B=2$ and
\textit{iii}) $\xi_0^{{}}/B=-4$. } \label{fig:CoPVg}
\end{figure}

For the antiparallel geometry the same angular dependence is found
for all three regimes with $G_{AP}^{(j)}\propto \sin^2\phi$.
Fig.~\ref{fig:CoPVg} shows the conductance for the parallel
geometry for the three different regimes. It is seen how the
destructive interference only occurs in regime \textit{i}), i.e.
when the levels are split on each side of the chemical potentials
and the interactions are not too strong. The angular dependence
for the two other regimes depends on which level is closest to the
chemical potential of the leads.

\section{Discussion and summary}

\label{sec:disc}

For the non-interacting case we found the exact linear response
conductance. For the parallel configuration we saw that the
angular dependence could be understood in terms of an interference
effect occurring due to the spin-splitting of the dot. For
$\phi=0$ the spin bases of the dot and the leads are identical,
and the electrons can only pass through the dot via the
spin-$\uparrow$ level. For a non-collinear magnetic field the dot
spin eigenstates are superpositions of the lead spin states, which
gives two possible paths through the dot. This leads to
destructive interference at $\phi=\pi/2$, and the resonance
sharpens as the tunneling coupling increases. This result is in
agreement with the cotunneling result in Eq.~\eqref{Jparr} setting
$U=0$. Interestingly, the destructive interference is destroyed by
interactions, giving rise to an interaction induced enhancement of
the conductance. Thus the conductance goes from being proportional
to $\cos^2\phi$ in the non-interacting case to being proportional
to $\cos^4(\phi/2)$ for the strongly interacting case. However, if
the leads are not fully polarized the angular dependence persists,
but is gradually washed out for decreasing polarization.

For the antiparallel geometry the situation is very different. In
the non-interacting case (and symmetric coupling) the conductance
is proportional to $\sin^2\phi$ for all values of $B/\Gamma^0$.
This is simply understood as a spin-valve effect, giving zero
conductance at $\phi=0$ and $\phi=\pi$, whereas for non-collinear
$B$-field the spin can flip when it passes the dot. Furthermore,
in this case there is a constructive interference at $\phi=\pi/2$.
Now, with interactions on the dot one spin channel becomes
partially blocked, but the angular dependence remains the same
even in the limit of large $U$. This is thus in stark contrast to
the above mentioned cross-over observed for the parallel
configuration.

We have thus focused on two cases: 1) strong tunneling and no
interaction, and 2) weak tunneling with interaction. The
cross-over between these two regimes is an interesting and
challenging issue, because a formalism that captures both
coherence and correlations on equal footing is needed.

\appendix

\section{Sequential tunneling limit for parallel magnetization}

\label{sec:seq}

In the main part, we have discussed the effects of higher order
tunneling where interference effects take place. For completeness,
we here give the results for the sequential tunneling limit for
parallel magnetization.

In the parallel case, we can use the conductance formula derived
by Meir and Wingreen\cite{meir92} for proportional coupling if we
also assume that $P=P_{L}=P_{R}$. With these assumptions, we have
\begin{equation}
G=\frac{ie^{2}}{h}\int d\omega \,\left( -\frac{dn_{F}^{{}}(\omega
)}{d\omega }\right) \mathrm{Tr}\left[ \mathbf{\Gamma }\left(
\mathbf{G}^{R}(\omega )-\mathbf{G}^{A}(\omega )\right) \right] .
\label{Gprop}
\end{equation}
where
\begin{equation*}
\mathbf{\Gamma }=\frac{\Gamma _{L}^{0}\Gamma _{R}^{0}}{\Gamma
_{L}^{0}+\Gamma _{R}^{0}}\left(
\begin{array}{cc}
1+P\cos \phi & P\sin \phi \\
P\sin \phi & 1-P\cos \phi
\end{array}
\right) .
\end{equation*}

To lowest order in the tunneling, which is the so-called sequential
tunneling case, the Green's functions in Eq.~\eqref{Gprop} are given by
\begin{equation}
i(\mathbf{G}^{R}-\mathbf{G}^{A})\!=\!\left(
\begin{array}{cc}
n_{\downarrow }^{{}}\delta _{\uparrow }^{U}+(1-n_{\downarrow }^{{}})\delta
_{\uparrow }^{0} & 0 \\
0 & n_{\uparrow }^{{}}\delta _{\downarrow }^{U}+(1-n_{\uparrow }^{{}})\delta
_{\downarrow }^{0}
\end{array}
\right) ,
\end{equation}
where
\begin{equation}
\delta _{\sigma }^{A}=2\pi \delta (\omega -\varepsilon_{\sigma
}^{{}}-A),
\end{equation}
and
\begin{equation}
n_{\uparrow ,\downarrow }=\frac{e^{-\beta (2\xi_{0}+U)}+e^{-\beta
(\xi _{0}\pm B)}}{1+e^{-\beta (2\xi _{0}+U)}+2\cosh (\beta
B)e^{-\beta \xi _{0}}},
\end{equation}
where $\beta $ is the inverse temperature. Thus for the parallel
case, the conductance becomes a sum of four peaks:
\begin{eqnarray}
\frac{G}{h/e^{2}}\frac{\Gamma_0}{\Gamma^0_L\Gamma_R^0} &=&\left(
h(\xi_0^{{}} +B)\left[ 1-n_{\downarrow }^{{}}\right]
+h(\xi_0^{{}} +B+U)n_{\downarrow }^{{}}\right)  \notag \\
&&\times \left( 1+P\cos \phi \right)  \notag \\
&&+\left( h(\xi_0^{{}} -B)\left[ 1-n_{\uparrow }^{{}}\right]
+h(\xi_0^{{}}
-B+U)n_{\uparrow }^{{}}\right)  \notag \\
&&\times \left( 1-P\cos \phi \right) ,
\end{eqnarray}
where $h(\omega )=\left( -\frac{dn_{F}^{{}}(\omega )}{d\omega
}\right).$ This agrees with the non-interacting result for $U=0$
and $\Gamma^0\rightarrow 0$.

For the antiparallel geometry it is more complicated to obtain a
simple result because the current formula in this case contains
the lesser Green's function.\cite{meir92} The full rate equation
results have been given in Ref.~\onlinecite{braun04}.

\end{document}